\documentclass{iau} 
\usepackage{graphicx}
\usepackage{multirow}

\title[Stability of the fine-structure constant with the ELTs] 
{Fundamental physics constraints from testing the stability of the fine-structure constant with the ELTs
}

\author[A. C. O. Leite, C. J. A. P. Martins, P. Molaro,  C. S. Alves, T. A. Silva]   
{A. C. O. Leite$^{1,2,3}$, C. J. A. P. Martins$^{1,2}$, P. Molaro$^{4}$,  C. S. Alves$^{2,5}$, T. A. Silva$^{2,3}$}
\affiliation{email: {\tt ana.leite@astro.up.pt} \\$^1$Instituto de Astrof\'{i}sica e Ci\^{e}ncias do Espa\c{c}o, Rua das Estrelas, 4150-762 Porto, Portugal \\ [\affilskip]
$^{2}$Centro de Astrof\'{i}sica, Universidade do Porto, Rua das Estrelas, 4150-762 Porto, Portugal\\[\affilskip]
$^3$Faculdade de Ci\^{e}ncias da Universidade do Porto, Rua do Campo Alegre, 4150-007 Porto, Portugal \\  [\affilskip]
$^4$ INAF-Osservatorio Astronomico di Trieste, Via G.B. Tiepolo 11, I-34143 Trieste, Italy \\ [\affilskip]
$^5$ Department of Mathematics, Imperial College London, London SW7 2AZ, United Kingdom }
\pubyear{2018}
\volume{347}  
\jname{Early Science with ELTs}
\editors{A.C. Editor, B.D. Editor \& C.E. Editor, eds.}
\begin{document}

\maketitle

\begin{abstract}
The increased collecting area of the ELTs will bring fainter high-z targets within the reach of high-resolution ultra-stable spectrographs, thus enabling a new generation of precision consistency tests, including tests of the stability of nature's fundamental couplings. For example, the stability of the fine-structure constant can be tested by looking at metal absorption lines produced by the intervening clouds along the line of sight of distant quasars. 

In this contribution we discuss the performance that can be expected from the ELTs in testing the stability of the fine-structure constant, based on the early ESPRESSO observations, and some comparative forecasts of the impact of these measurements for representative models of fundamental physics and cosmology.
\end{abstract}
\firstsection 
\section{Introduction}
Quasar (QSO) absorption spectra are powerful laboratories to test the variation of fundamental constants. Absorption lines produced by the intervening clouds along the line of sight of the QSO give access to physical information on the atoms present in the cloud, and this means that they give access to physics at different cosmological times and places. Different energy levels of some atomic transitions in these clouds are sensitive to variations of the fine-structure constant, $\alpha$, and this sensitivity varies according to the different atomic structures. If the physics in the cloud where the absorption line is produced was different, then one would get position shifts compared with the laboratory wavelengths corrected for the redshift. We fit as many sensitive lines as possible to break degeneracies and to increase the statistical signal of the measurement.

The best measurements of $\alpha$ in QSO spectra existing today were obtained with two high-resolution spectrographs: UVES on the VLT; and HIRES, at the Keck Observatory.
A data set of 293 archival measurements from these two instruments show a spatial variation of $\alpha$ at more than four-sigma level of statistical significance, at the ppm level (\cite{webb2011}). There is no identified systematic effect that is able to fully explain it. A detailed analysis of the data reduction, dipole computation and search for systematics is described in King (2011).
After  this study some systematics in the wavelength calibration of data collected with high-resolution, slit spectrographs  were identified.
Rahmani et al. (2013) found long-range wavelength distortions in the VLT/UVES instrument and other following studies corroborated these findings and showed that these distortions vary over time (\cite{bagdonaite2014,evans2014,songaila2014,whitmore2015}).
          These effects could imply significant errors within the previous data set of measurements, weakening the evidence for the claim of the dipole variations in $\alpha$. 

ESPRESSO is a new high-resolution, ultra-stable spectrograph that started observations in September 2018. Being fibre-fed will allow for precise $\alpha$ measurements with control of the previous identified systematics. The instrument is installed at the VLT and offers the possibility of observing individually with one Unit Telescope, or combining the light of the four telescopes allowing fainter targets to be reached (\cite{espresso}). 

\section{Wavelength coverage}

In the context of choosing the best targets to test the stability of $\alpha$ with ESPRESSO we defined in previous work (\cite{leiteGTO}) what an 'ideal' target to achieve higher constrains on $\alpha$, should be. Summarizing, we considered the $\sim$300 reported measurements and took into account the transitions not observable with ESPRESSO because of the shorter wavelength coverage compared to the previous two spectrographs.

 In order to have a sensitive measurement on $\alpha$ with a better control of systematics one should ensure that this target has transitions that are sensitive to an $\alpha$ variation, i.e., it should have at least one transition that shifts to the blue, one to the red, and one that is not sensitive to a variation. This demands that the sensitivity coefficient $\Delta q$ for a given system is large, ensuring that we are able to control the fit for $\alpha$, and constrain redshift and intrinsic velocity structure of the cloud simultaneously. We chose targets with $\Delta q>2000$ and a reported uncertainty of $\sigma_{\Delta\alpha/\alpha}<5$\,ppm. This last criteria comes from the fact that simple spectra should have already produced measurements with statistically lower uncertainties.
 
 \begin{figure}[b]
 \vspace*{-3 mm}
\begin{center}
\includegraphics[width=2.7in]{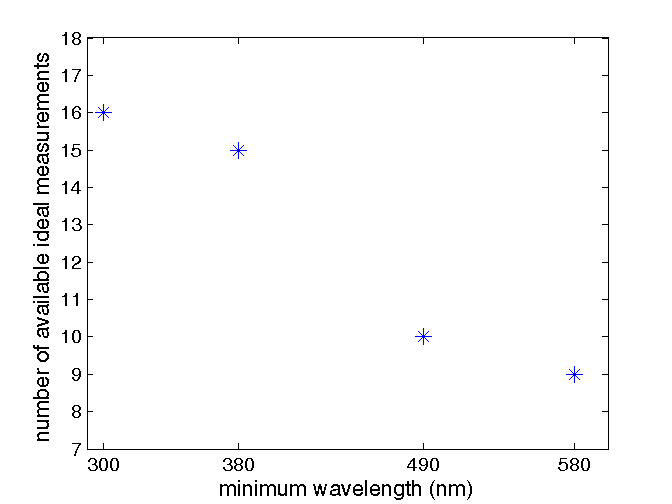} 
 \vspace*{-0.1 cm}
 \caption{The number of ideal targets (defined in the text) depends on different wavelength cutoffs at the blue end of the spectrum. The maximum wavelength is 1000 nm in all cases.}
   \label{fig1}
\end{center}
\end{figure}
 
ELT-HIRES planned for the ELT should be the next generation high-resolution, ultra-stable spectrograph after ESPRESSO (\cite{elt}). 
 Testing the stability of fundamental constants is one of the design drivers of HIRES. We should expect higher sensitivity on measurements just from the increased collecting area and possible much better control over systematics.
The wavelength coverage of ELT-HIRES is a critical point to this science goal. Although the spectral coverage in the red will certainly be at least the same as previous spectrographs, the wavelength cutoff in the blue remains undefined, and certainly has an impact on the number of available targets that can provide an  $\alpha$ measurement. The specific final blue cutoff wavelength will limit the number of available lines that can be used to measure $\alpha$, and more critical, limit the  sensitivity of measurements.
There are very few blue shifter transitions, the most common one being Fe\,II\,1608\,\AA. If the wavelength coverage in the clue starts at 300~nm then we are able to measure this blue shifter transition in targets with $z>0.87$.  If instead we consider the other extreme, with the wavelength coverage in the blue starting at 580~nm, then only targets with $z>2.6$ can, in principle, offer a good sensitivity.
 
Following the exercise that was done for ESPRESSO,  we defined  the number of 'ideal' targets for $\alpha$ for different spectrograph wavelength ranges. We present in Fig. \ref{fig1} the number of ideal targets that have the required characteristics of a good target for various differenet wavelength cutoffs in the blue wavelength coverage of different considered blue ends for a spectrograph (and with the wavelength coverage in the red fixed at 1000\,nm). For comparison ESPRESSO can observe 14 'ideal' targets. A wavelength cutoff higher than 380\,nm eliminates half the sample of 'ideal' targets, limiting the impact that future ELTs can have on testing fundamental constants.

\section{Forecasts on Cosmology -  Bekenstein-type models}

These types of measurements are important to test the stability of $\alpha$ but even a non-detection of variation  can be used to constrain dark energy. For the forecast on the following sections, we assumed the 14 targets of ESPRESSO GTO with the expected baseline uncertainties of $\sigma_\alpha=0.6$ ppm and assuming  the same  wavelength coverage of ESPRESSO for the case of ELT-HIRES, we expect the baseline uncertainties of $\sigma_\alpha=0.1$\,ppm.

The Bekenstein-Sandvik-Barrow-Magueijo model assumes a variation of $\alpha$ that comes  from the variation of the electric charge (\cite{bekenstein}). This  model can be seen as a $\Lambda$CDM-like model with an additional dynamical degree of freedom,whose dynamics is such that it leads to the aforementioned variations without having a significant impact on the Universe's dynamics. 

The constraints on these kinds of models from current astrophysical and cosmological data as well as detailed forecasts for ESPRESSO and ELT-HIRES are presented in Leite \& Martins (2016). One important variable is the coupling $\zeta_{\alpha}$, a free parameter of the model that gives magnitude to the allowed variation on $\alpha$. Assuming the baseline uncertainties on $\alpha$ expected for ESPRESSO and ELT-HIRES as described above, one gets the results presented on the first columns of Table \ref{tab3}. We expect  improvements relative to the current constraints on the coupling $\zeta_\alpha$ by a factor of around 5 for ESPRESSO and 50 for ELT-HIRES.

\textbf{Olive \& Pospelov extension} -- An extension to this type of model was discussed by Olive \& Pospelov (2002), allowing for different couplings to the dark matter and dark energy sectors and the behaviour of $\alpha$ depending on both of them. Constraints on these kinds of models from current astrophysical and cosmological data as well as detailed forecasts are presented in Alves et al. (2018). The important variables in this case are the couplings $\zeta_{m}$ and $\zeta_{\Lambda}$,  free parameters of the model that give magnitude to the allowed variation on $\alpha$.
Assuming the baseline uncertainties on $\alpha$ expected for  ESPRESSO and ELT-HIRES as described above, one gets the results in the last two columns of Table \ref{tab3} for the two free parameters of this model. We compare it with the constraints of all the existing measurements of $\alpha$ ($\sim300$) and we can see that for the 14 measurements of the ESPRESSO GTO, we get comparable constraints, but for ELT-HIRES, the improvements are evident.

\begin{table}[h]
\centering
\caption{\small\label{tab3} One sigma uncertainties (in ppm) on the couplings $\zeta_\alpha$ for the Bekeinstein model, and on the couplings $\zeta_m$ and $\zeta_\Lambda$ for the Olive \& Pospelov (2002) extension model. Constraints come from assuming existing $\alpha$ data, and the corresponding expected forecasts for the ESPRESSO Fundamental Physics GTO target list and  for ELT-HIRES}. 
 
{\scriptsize
\begin{tabular}{|l|c|c c|}
\hline
 & Bekenstein & \multicolumn{2}{ |c| }{Olive \& Pospelov}\\
\hline
Data set& $\zeta_\alpha$ & $\zeta_m$ & $\zeta_\Lambda$ \\
\hline
Current constraints& $1.7$ & $0.72$ & $1.84$ \\ 
\hline
ESPRESSO Baseline & $0.21$& $0.73$ & $4.36$\\ 
\hline
ELT-HIRES Baseline& $0.02$ &$0.12$ & $0.73$ \\
\hline
\end{tabular}
}
\end{table}

\section*{Acknowledgments}
This work was financed by FEDER - Fundo Europeu de Desenvolvimento
Regional funds through the COMPETE 2020 - Operacional Programme for
Competitiveness and Internationalisation (POCI), and by Portuguese
funds through FCT - Funda\c{c}\~{a}o para a Ci\^{e}ncia e a Tecnologia in
the framework of the project POCI-01-0145-FEDER-028987. ACL is supported by an FCT
fellowship (SFRH/BD/113746/2015), under the FCT PD Program PhD::SPACE (PD/00040/2012).

\begin{discussion}
\discuss{P. Noterdaeme}{How is the number of 'good' targets for $\Delta\mu/\mu$, decreasing with decreasing wavelegth coverage? In other words: How many $\Delta\mu/\mu$ targets are there for GMT, TMT, ELT?}
\discuss{A.C.O.Leite}{We didn't specifically study the different ELTs cases. But knowing the rest frame wavelength of $H_2$ transitions used typically in the optical to perform   $\Delta\mu/\mu$ measurements, we computed  for ESPRESSO with a blue cut of at 380nm, we can barely perform 1 measurement, for a single target at $z=3.025$. For a blue cut off above that there are no targets for $\mu$ measurements that can be done using  $H_2$.}
\discuss{Dmetrios Matsakis}{If you detect a variation of $\alpha$, can you reconcile it with atomic clock data on Earth?}
\discuss{A.C.O.Leite}{The current atomic clock measure $\dot{\alpha}/\alpha$ locally ($z=0$)  with better precision than QSO absorption spectra. 
They find that the current drift per year of $\alpha$ is consistent with 0 at the $10^{-17}$ level (Rosenband et al. 2008). That being said, the QSO targets measure $\alpha$ at higher redshift, and in different cosmological environments. There are different models that allow a variation of $\alpha$ that evolves with time, and others that allow for local dependencies such as density. 
} 
\end{discussion}
\end{document}